\documentclass{aa}  
\usepackage[utf8]{inputenc}
\usepackage{natbib}
\usepackage{hyperref}
\usepackage{threeparttable}
\usepackage{gensymb}
\usepackage{tabularx}
\usepackage{booktabs}
\usepackage{amsmath}
\usepackage{amssymb}
\usepackage{version}
\usepackage{graphicx}
\usepackage{pdfpages}
\usepackage{booktabs}
\usepackage{caption}
\usepackage{multirow}
\usepackage{comment}
\usepackage{graphicx}
\usepackage{appendix}
\usepackage{nicematrix}
\usepackage{siunitx}
\usepackage{lscape} 
\usepackage[font={footnotesize}]{caption}
\usepackage{pdflscape}
\usepackage[margins]{trackchanges}
\usepackage{textcomp}
\newcommand{\ha}{H$\alpha$}

\usepackage{txfonts}
\hypersetup{
    colorlinks=true,
    linkcolor=blue,
    filecolor=magenta,      
    urlcolor=cyan,
    citecolor=blue,
}
\begin{document} 

\title{$H\beta$ spectroscopy of the high-inclination black hole transient Swift J1357.2-0933 during quiescence
 }
   
\author{A. Anitra,\inst{1} D. Mata Sánchez,\inst{2,3} T. Muñoz-Darias,\inst{2,3} T. Di Salvo,\inst{1}  R. Iaria,\inst{1} C. Miceli,\inst{1,4} M. Armas Padilla,\inst{2,3} J. Casares,\inst{2,3} J. M. Corral-Santana\inst{5} }

   \institute{Dipartimento di Scienze Fisiche ed Astronomiche, Università di Palermo, via Archirafi 36, 90123 Palermo, Italy
   \and 
   Instituto de Astrofísica de Canarias, 38205 La Laguna, Tenerife, Spain
   \and
   Departamento de astrofísica, Univ. de La Laguna, E-38206 La Laguna, Tenerife, Spain
   \and
INAF/IASF Palermo, via Ugo La Malfa 153, I-90146 Palermo, Italy 
    \and
European Southern Observatory (ESO), Alonso de Cordova 3107, Vitacura, Casilla 19001, Santiago, Chile}

 
\abstract{
Swift J1357.2-0933 is a transient low-mass X-ray binary hosting a stellar-mass black hole. The source exhibits optical dips and very broad emission lines during both outburst and quiescence, which are thought to be the result of a high orbital inclination. We present phase-resolved spectroscopy obtained with the 10.4m Gran Telescopio Canarias (GTC). 
The spectra focus on the $\rm{H}\beta$ spectral region during X-ray quiescence. The emission line is exceptionally broad (full width at half maximum, FWHM > 4000 \AA), in agreement with previous studies focused on $\rm{H}\alpha$. A two-Gaussian fit to the prominent double-peaked profile reveals a periodic variability in the centroid position of the line. We also produced a diagnostic diagram aimed at constraining additional orbital parameters. Together, they allow us to independently confirm the orbital period of the system using a new dataset obtained five years after the previous outburst. However, our estimates for both the systemic velocity and the radial velocity semi-amplitude of the black hole reveal larger values than those found in previous studies. We argue that this could be explained by the precession of the disc and the presence of a hotspot.\\
We found evidence of a narrow inner core in the double-peaked H$\beta$ emission profile. We studied its evolution across the orbit, finding that it is likely to result from the occultation of inner material by the outer rim bulge, further supporting the high orbital inclination hypothesis.}
\titlerunning{Swift J1357.2-0933: The high-inclination black hole X-ray transient}
   \authorrunning{A.Anitra}
   \maketitle
%
\section{Introduction}\label{sec_1}
Low-mass X-ray binaries (LMXBs) are composed of a compact object, either a neutron star or a black hole (BH), and a low-mass companion star (<$1M_\mathrm{\odot}$). In such systems, the accretion of mass from the companion onto the compact object occurs via an accretion disc and ultimately powers their extreme phenomenology.
The vast majority of BHs in LMXBs are transient, spending most of their lifetimes in a quiescent state characterised by an X-ray luminosity of about 10$^{30-34} \rm{erg \, s^{-1}}$ \citep{2014_armasB}, but occasionally undergo dramatic outbursts during which their emission increases by several orders of magnitude.  

Swift J1357.2-0933 (hereafter J1357) is a LMXB X-ray transient discovered during its outburst in 2011, with a lower limit to the distance of d > 2.29 kpc that, combined with the high Galactic latitude, places it above the Galactic thick disc (see \citealt{Daniel_2015}).
Despite the lack of detection of its companion star during quiescence, estimations of the compact object mass based on the full width at half maximum (FWHM)--K2 correlation \citep{Casares_2015} suggest that J1357 harbours one of the most massive stellar-mass BHs known in the Galaxy ($\rm{M_{BH}} > 9.3 \,M_{\odot}$, as reported by \citealt{Daniel_2015}).

The peak X-ray luminosity estimated during the outburst is about $\rm{L}_{x}^{peak} \sim 10^{35}(\rm{D/1.5 \,kpc})^{2}$ ${\rm{erg \, s^{-1}}}$ \citep{monse_2013}, placing the source among the very faint X-ray binary transients (VFXTs) \citep{2006_Wijnands}.
The system's orbital period was measured to be $2.8\pm 0.3\, \rm{h}$ from the variability on the double-peaked H$\alpha$ emission line, first during outburst \citep{Corral_santana_2013} and subsequently during quiescence \citep{Daniel_2015}. A refined value $2.5673\pm 0.0006\, \rm{h}$ was later obtained by studying the modulation of the H$\alpha$ line profile trough over 14 months \citep{Casares_2022}

One of the most puzzling properties of J1357 is its dipping activity: the optical light curve shows periodic dips in its evolution during the outburst, while the X-ray light curve does not \citep{Armas_2014}. The origin of this behaviour is still debated. For instance, \citet{Corral_santana_2013} associated this behaviour with the presence of an obscuring toroidal structure in the inner accretion disc that moves towards the outermost regions during the outburst decay. 
In their multi-wavelength photometric analysis \cite{paice_2019} found that the source becomes bluer during the dips, and tried to give an updated model that describes the system geometry, proposing that the obscured emission mainly arises from the base of the jet. This was later supported by \citet{2019_jimenez-ibarra}, who also performed dip-resolved spectroscopy, finding that the optical dips were associated with blue-shifted H and He II absorptions. This behaviour, also observed by \cite{Charles_2019}, suggests that the dips are related to wind-type outflows.

Even after much debate on the geometry of the system, the dipping activity was found to lead inevitably to a high inclination angle of the system, with most of the literature on the system agreeing on an inclination angle greater than 70$\degree$ \citep{Corral_santana_2013}. For instance, \citet{Torres_2015} presented time-resolved optical spectroscopy, where the authors found evidence of time-variable deep absorption features in the double-peaked profile, supporting the hypothesis that the source is seen at a high inclination.
Typically, a high-inclination system exhibits eclipses in its X-ray light curve (e.g. MAXI J1659-152, \citealt{2012_Kennea}, \citealt{2013_Kuulkers}; 4U 1822-371 \citealt{anitra_2021}, \citealt{2013_Iaria}), and the absence of such eclipsing activity in this particular system may appear at odds with the hypothesis of high inclination. However, \citet{Corral_santana_2013} justified the lack of eclipses assuming a high orbital inclination >70$\degree$ and a very low mass ratio, which implies a donor star radius comparable to, or even smaller than, the outer rim of the disc.

Observations collected by the Swift X-ray telescope revealed an equivalent hydrogen column density ($\rm{N_H}$) of $(1.2 \pm 0.7)\times 10^{20} \, \rm{cm^{-2}}$ \citep{Torres_2015}. Such a low $\rm{N_H}$ implies a low optical extinction, providing us with a unique opportunity to obtain phase-resolved spectra of the system at blue optical bands with sufficiently high S/N, a rare occurrence in BH transients.

In this work we present the first quiescence spectroscopy of J1357 focused on the hydrogen H$\beta$ emission line and covering a full orbit. We analyse its periodic evolution and compare the results with those previously reported using the  $\rm{H}\alpha$ line during both quiescence and outburst periods \citep{Daniel_2015,Corral_santana_2013}.

\section{Observations}
We analysed eight observations collected by the Optical System for Imaging and low-intermediate-Resolution Integrated Spectroscopy (OSIRIS) at the 10.4 metre Gran Telescopio Canarias (GTC) at the Observatorio del Roque de los Muchachos (ORM) on the island of La Palma (Spain).
During the campaign we employed the R2000B grism with a 0.8'' slit, which allowed us to analyse the spectral features in the wavelength range  $3950-5700\,$\AA\, with a spectral resolution of R=1903 and dispersion of D = 0.86 \AA/pix. The resolution is estimated by measuring the FWHM of the skylines in the background spectra, while the dispersion value is measured at $\lambda_{c}$ = 4755 \AA, as reported in the  OSIRIS user manual.
The eight observations were acquired consecutively, during a quiescent epoch, on the night of 2016 March 05 between 03:32:11.4 and 06:19:51, each with an exposure time of 1235 s, for a total exposure of 2.75 h, thus covering a full orbit of the system.  

The data were reduced using \textsc{IRAF}\footnote{IRAF is distributed by the National Optical Astronomy Observatories, operated by the Association of Universities for Research in Astronomy, Inc., under contract with the National Science Foundation.} standard routines, applying the bias and flat-field correction and using the HgAr, Xe, and Ne arcs lamps available for our configuration to calibrate the pixel-to-wavelength solution.
The spectra were corrected from cosmic rays using \textsc{L.A.Cosmic} task  \citep{2012_lacosmic}, and extracted from the two-dimensional images using the optimal extraction technique \citep{1998_Naylor}.

\section{Spectral analysis}\label{sec_3}
The analysis reported in this section was performed by using both \textsc{MOLLY} software,\footnote{The MOLLY software was developed by T. R. Marsh, and is available at the following link: https://cygnus.astro.warwick.ac.uk/phsaap/software/} and tailored \textsc{python} routines based on the \textsc{astropy} package \citep{astropy}.
We first normalised the spectra by dividing them by a third-order polynomial in order to model the continuum component of the spectrum. 
The wavelength coverage of our spectra allows us to access the H$\beta$ line region with a sufficiently high signal-to-noise ratio ($\rm{S/N}\sim\rm{3}$, measured in the nearby continuum). The H$\gamma$ line was also present, but the lower $\rm{S/N}$ (barely $\sim$ 1.3), mainly due to the lower response at the edge of the chip, together with a line intensity reaching barely half that of H$\beta$, prevented us from extracting a reliable line profile from the individual spectra. No other features are present in the spectra. 
We analysed for the first time the blue part of the spectrum, never studied in quiescence. We focused on the double-peaked H$\beta$ line, a signature profile of the gas orbiting in a Keplerian accretion disc.

\subsection{H$\beta$ radial velocity\label{radial velocity}}
\begin{figure}[!ht]
\centering
    \includegraphics[width=0.5\textwidth]{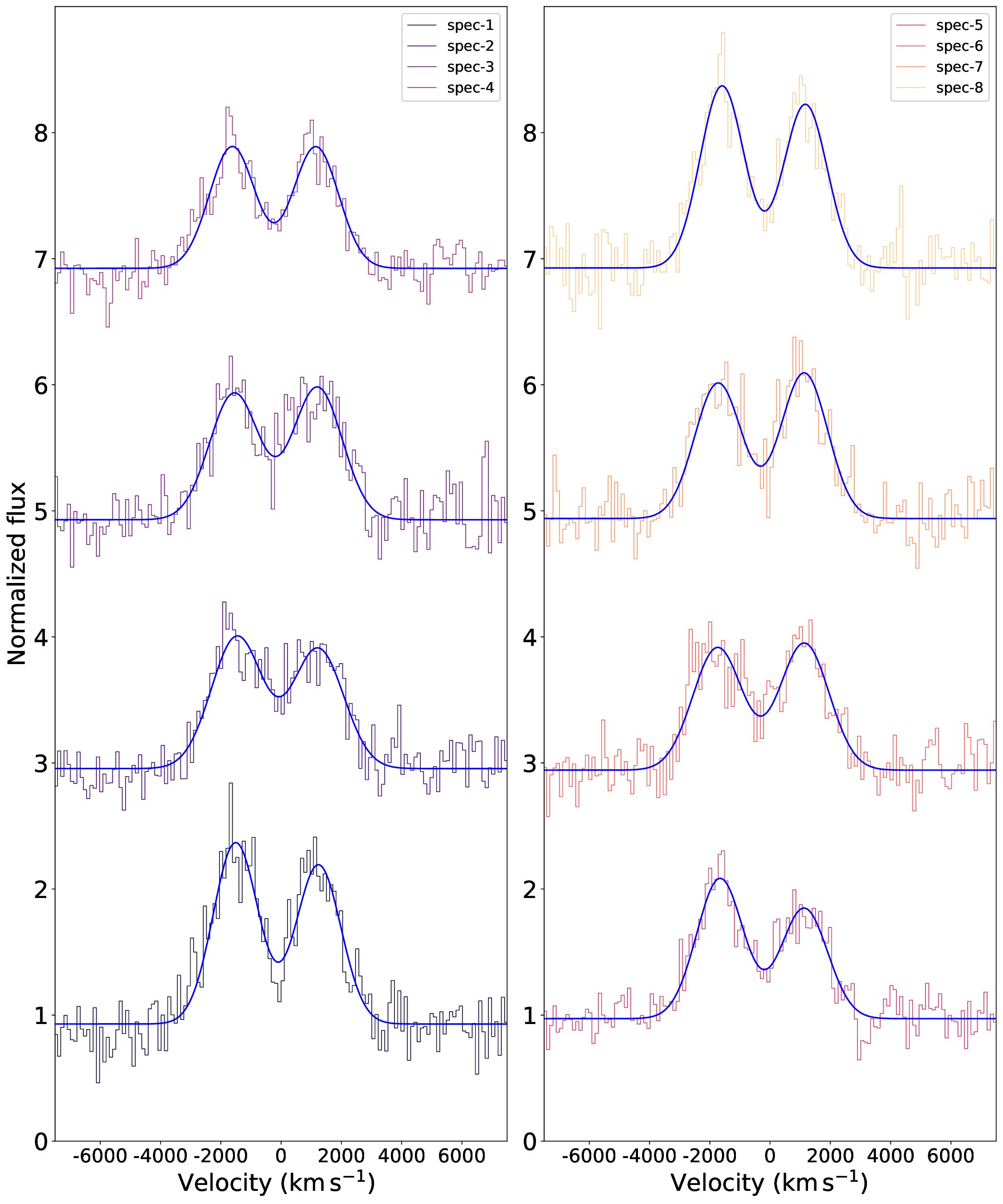}
    \caption{Observed spectra and the two-Gaussian model (solid blue line) of all the observations plotted vs the velocity Doppler shift related to the H$\beta$ rest-frame wavelength (4861 \AA). The spectra are binned by a factor of 2 and offset by a constant for visualisation purposes.}
    \label{fig:2fit}
\end{figure}

We fit every spectrum with a double-peaked model compound of two Gaussian lines using the Python \textsc{curve fit} routine from the \textsc{scipy} package \citep{2020SciPy-NMeth}. The free parameters controlling the fit are the intensity of each of the peaks; the FWHM, assumed to be the same for both Gaussians; and the shift of the centroid between the peaks with respect to the rest-frame wavelength, denoted offset.
All the best-fit parameters with the associated uncertainties are reported in Table \ref{2gaussfit}, while in Fig. \ref{fig:2fit} we show the fits of each individual spectrum.

We find periodic variability in the individual offsets of each observation, which led us to perform a non-linear fit to the data using the following function: 
\begin{equation}\label{sin}
 f(T)=\gamma+K_\mathrm{1}\sin{(2\pi((T-T_{0})/P-\phi_{0}-0.5))}.  
\end{equation}
Periodic radial velocity curves were previously observed in the H$\alpha$ line of other LMXBs, and in particular of J1357 \citep{Corral_santana_2013,Daniel_2015}. The interpretation is that it reflects the movement of the accretion disc centre of mass, which should therefore trace that of the compact object. 
Under the assumption of a perfectly symmetric disc, the average offset value across all of our observations  (i.e. $\gamma$) gives an estimate of the systemic velocity of J1357,   the parameter $K_\mathrm{1}$ is an estimate of the radial velocity semi-amplitude,    $\phi_{0}$ is an orbital phase shift correction conveniently defined, and the time of the first epoch of observations is $T_{0}= 2457452.64776595 \, \rm{HJD}$.

We observe a modulation with a $\rm{P}= 0.127 \pm 0.008\, \rm{d}$, consistent with previous determinations of the orbital period by \citet{Corral_santana_2013} and \citet{Daniel_2015}. \citet{Casares_2022} provide the most precise determination of the orbital period inferred for the source to date ($\rm{P}= 0.106969 \pm 0.000023\, \rm{d}$), based on the variability of the emission line over a number of orbits across different epochs. We note that our result is consistent only within 2.5$\sigma$ of their value. Our study is based on radial velocity measurements, which are most affected by perturbations due to the potential precession of the accretion disc (see Sections \ref{sec_3} and \ref{sec_4}); moreover,  it does not take account of the systematic effects due to our having only one orbit (which could lead to an underestimation of the error associated with the period), while the result of \cite{Casares_2022} is based on the study of a longer dataset covering several orbits. Together, they led us to adopt their more precise value and readjust the dataset by fixing this parameter.

The sinusoidal function fits the data closely, obtaining a chi-square over degrees of freedom (dof) of 0.75/5, which ensures a level of confidence (c.l.) of 97.5$\%$. Such a low $\rm{\chi^{2}/dof}$ suggests that the uncertainties on the original spectra are probably overestimated. Rescaling them to a factor of 0.86 produces a $\rm{\chi^{2}/dof=1.0}$.
We find the system is moving away from us at a systemic velocity of $\gamma=-220\pm$ 10 $\rm{km\,s^{-1}}$ and a radial velocity semi-amplitude of $K_\mathrm{1}= 91 \pm 16 \,\rm{km\,s^{-1}}$, as well as   $\phi_{0}=0.41 \pm 0.02$. 

\begin{figure}  
    \centering
    \includegraphics[width=.5\textwidth]{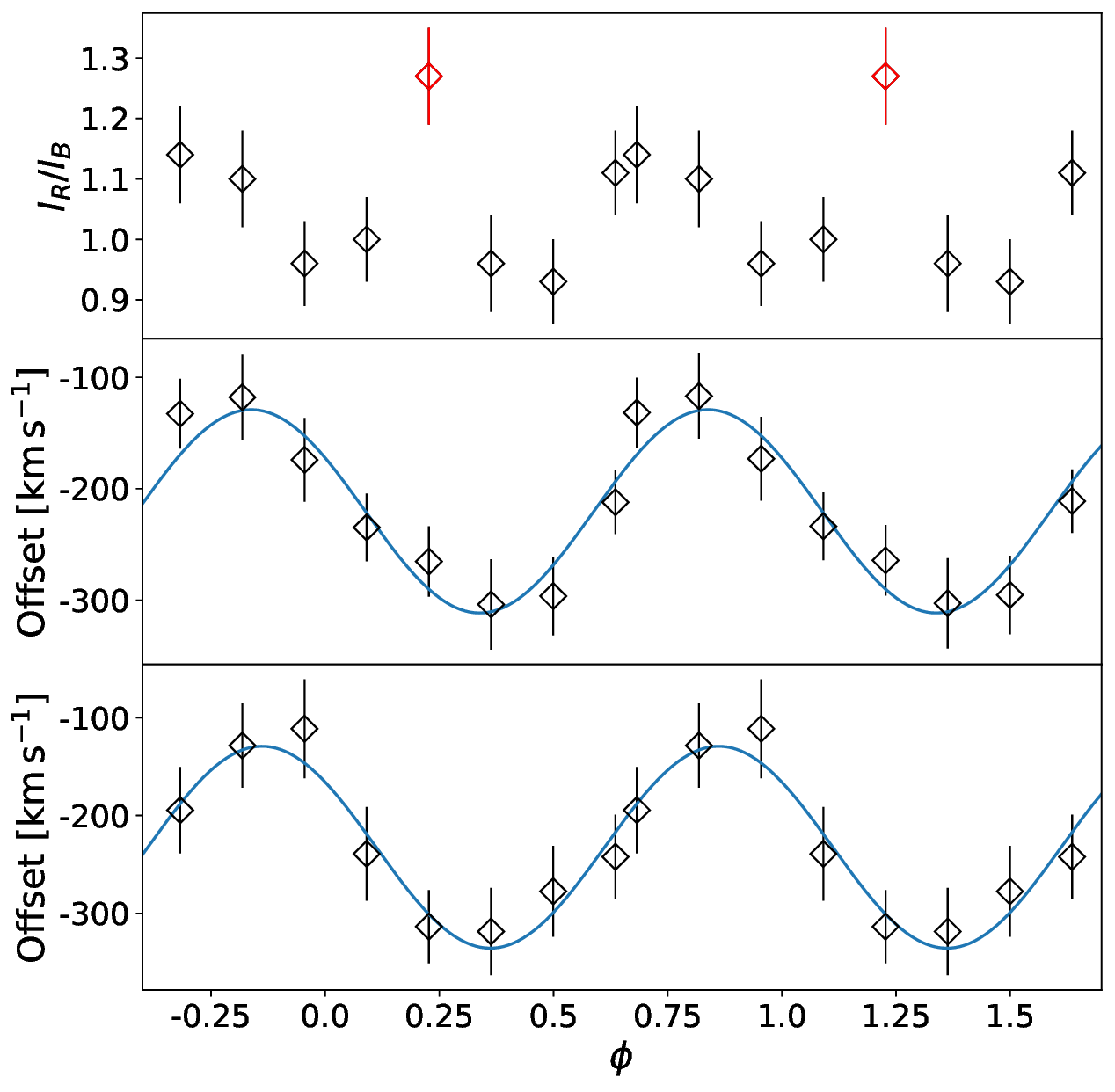}
        \caption{\label{variation_plot} Orbital evolution of H$\beta$ parameters obtained from the two-Gaussian model and a diagnostic diagram. They are all plotted against the orbital phase, and repeated over two orbits for visualisation purposes.\protect\\ 
        \textbf{Top panel}: Intensity ratio between the blue and red peak. The value deviating from the trend is in red. 
        \textbf{Middle panel}: Offset of the line centroid (black diamonds) and best fit from the double-peaked model (blue solid line).
        \textbf{Bottom panel}: Offset of the line wings centroid (black diamonds) and best fit from the diagnostic diagram (blue solid line), corresponding to a peak-to-peak separation of $a=3800 \, \rm{km s^{-1}}$.}
\end{figure}

Under this interpretation, and following the usual definition for zero phase (i.e. corresponding to inferior conjunction), we can calculate the orbital phase for any given epoch using the following equation: $\phi = (T - T_{0})/P - \phi_{0}$.
In Table \ref{2gaussfit} we report the relative orbital phase of each observation, under the above convention, and we plot our best fit in Fig. \ref{variation_plot} (middle panel).
\begin{table}\centering
 \caption{\label{2gaussfit} Best-fit values for the parameters of the two-Gaussian model described in the text. Uncertainties are at the 68\% c.l.}
 \begin{threeparttable}
  \resizebox{0.5\textwidth}{!}{\begin{minipage}{\textwidth}
\begin{tabular}{lccccccc}
\hline
Spec & phase & phase\tnote{$\dagger$} & offset ($\rm{km\, s^{-1}}$)  & FWHM ($\rm{km\, s^{-1}}$)      & $I_{R}/I_{B}$ &  $\chi^{2}/d.o.f.$ & T-value \\ \hline 
spec-1   &  -0.32 & -0.11 & -133 $\pm$ 31 & 1762 $\pm$ 64                     & 1.14 $\pm$ 0.08           & 1.40 & $ 0.64 \pm  0.02$
 \\
spec-2   & -0.18&  0.03& -118 $\pm$ 38     & 2012 $\pm$ 79                     & 1.10 $\pm$ 0.07       & 1.30  & $ 0.43 \pm  0.02$
 \\
spec-3  & -0.05 & 0.16 &-174 $\pm$ 38     & 1946 $\pm$ 78                     & 0.96$\pm$ 0.07               & 1.76 & $ 0.51 \pm  0.02$
 \\
spec-4  & 0.09  &0.30 & -235 $\pm$ 30      & 1794 $\pm$ 64                     & 1.00 $\pm$ 0.07             & 1.75 &  $ 0.62 \pm  0.02$
\\
spec-5 & 0.23  & 0.43& -265 $\pm$ 32      & 1835 $\pm$ 65                     & 1.27 $\pm$ 0.08             & 1.36 &$ 0.59 \pm  0.02$
   \\
spec-6  & 0.36  & 0.57& -304 $\pm$ 41     & 1950 $\pm$ 83                     & 0.96 $\pm$ 0.08                & 1.34 & $ 0.57 \pm  0.02$
 \\
spec-7 & 0.50  & 0.71& -296 $\pm$ 35      & 1845 $\pm$ 73 &                     0.93 $\pm$ 0.07   & 1.28 & $ 0.63 \pm  0.02$
     \\
spec-8 & 0.64 & 0.84&-212 $\pm$ 28      & 1724 $\pm$ 59                     & 1.11 $\pm$ 0.07        &1.29 & $ 0.67 \pm  0.02$ \\                    \hline            
\end{tabular}
\end{minipage}}
\begin{tablenotes}
\footnotesize
\item[$\dagger$] Phases relative to the absolute inferred by \cite{Casares_2022}. 
\end{tablenotes}

\end{threeparttable}
\end{table}

\begin{figure}  
    \centering
    \includegraphics[width=.5\textwidth]{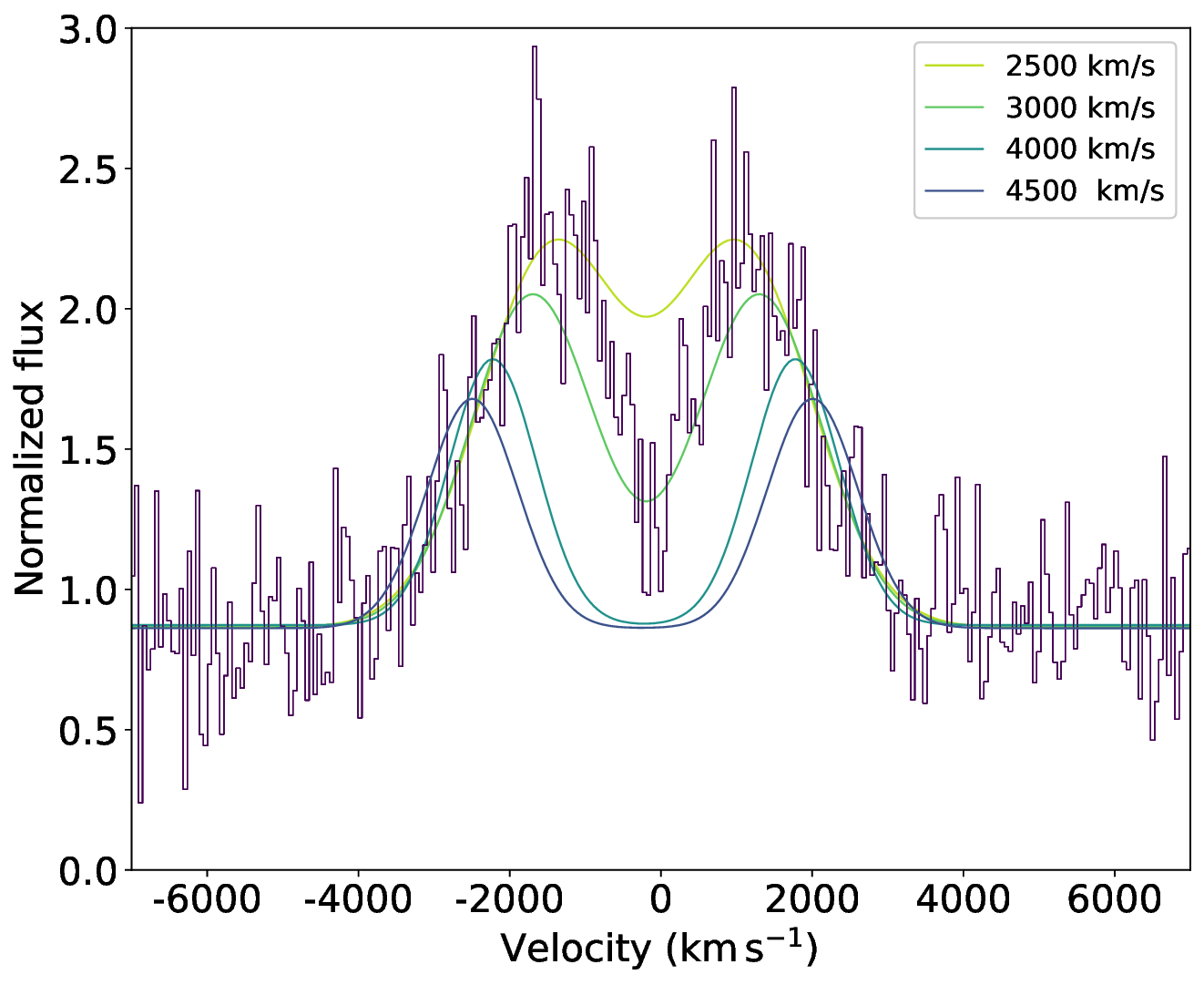}
        \caption{\label{prova_diagnositc} Visual example of the diagnostic diagram best-fit model to spec-1 for different peak-to-peak separation values.}
\end{figure}

\subsection{Diagnostic diagram \label{diagnostic}}
In our analysis we note a peculiar behaviour of the variation in the ratio of the intensity of the blue to the red peaks of the H$\beta$ profile.
The evolution of the intensity variation during the orbital period is shown in Fig. \ref{variation_plot} (top panel). There is an apparent periodic modulation of this ratio, similar to that observed previously in the H$\alpha$ line (see \citealt{Daniel_2015}), but the ratio associated with spectrum 5 (spec-5) deviates from the expected trend (a visual inspection confirms the blue peak is clearly more intense than the red peak). 

This phase-dependent asymmetry of the line profile can be due to different effects;   the precession of a disc or the contamination by the presence of a hot spot are the traditionally invoked explanations \citep{shafter_1986}.
Consequently, they might complicate the interpretation of the periodic variability described in the previous section: while the presence of a variable hot spot can bias the K$_{1}$ determination, a precessing disc might lead to variable $\gamma$ values when measured in different epochs.

\begin{figure}[!ht]
\centering
    \includegraphics[width=.5\textwidth]{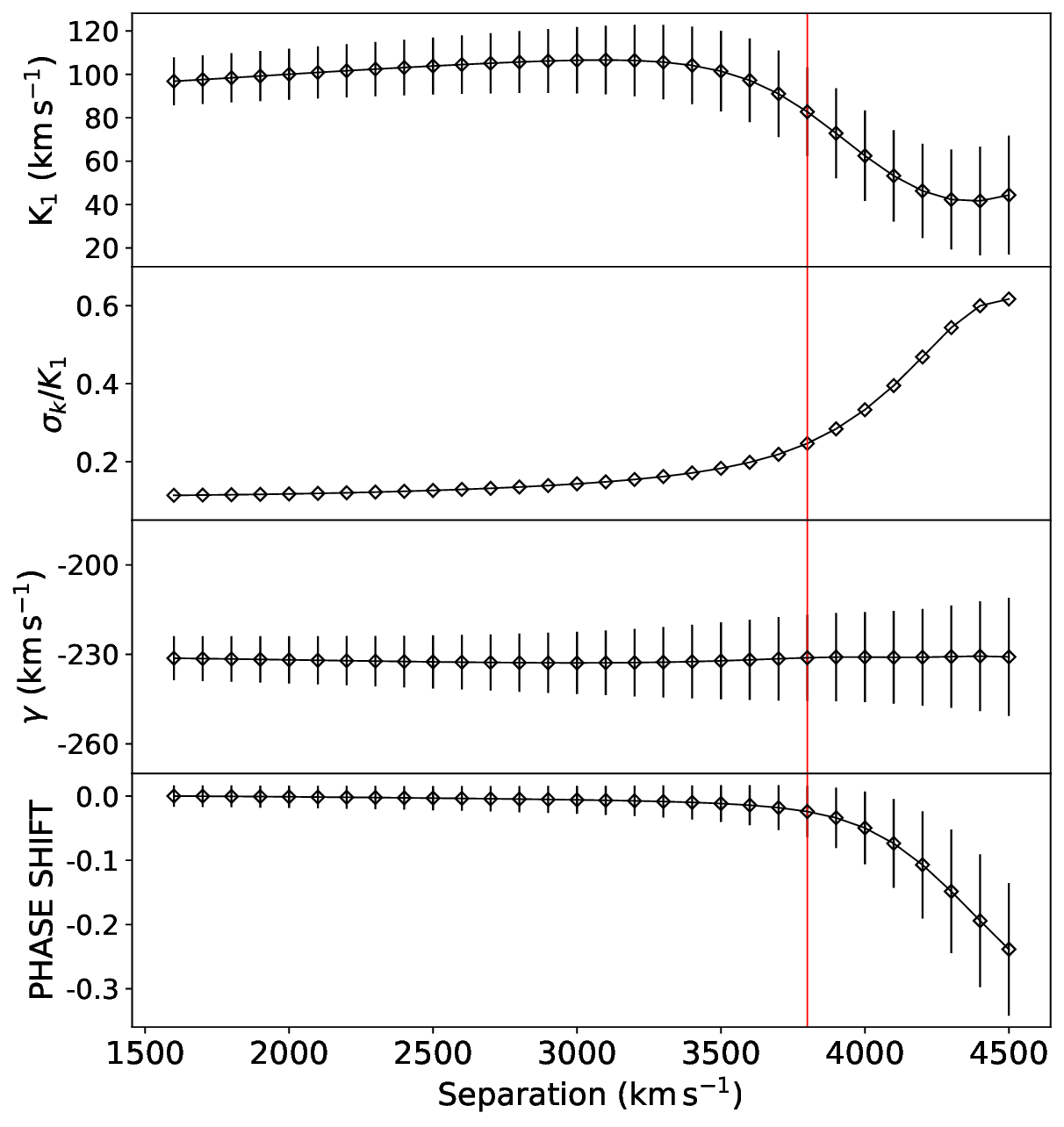}
            \caption{\label{diagnostic} Diagnostic diagram for  H$\beta$.\protect\\ Every parameter was calculated by using two Gaussian lines with the same FWHM. Shown from top to the bottom,  plotted vs the peak-to-peak separation, are  the semi-amplitude velocity $K_\mathrm{1}$, the ratio of $K_\mathrm{1}$ to its associated error, the systemic velocity $\gamma$, and the corresponding phase shift. The vertical red line represents the peak-to-peak separation ($a=3800 \,\rm{km\,s^{-1}}$) above which the measurements begin to be contaminated by noise in the continuum, as inferred from the plot in the second panel.}
\end{figure}

Consequently, in order to get a more reliable measure of the radial velocity variation and to avoid phase-dependent contamination, we decided to focus our analysis on the innermost regions of the disc accessible through optical spectroscopy.

An efficient method of investigating the variation in the radial velocity of the line wings was proposed by \cite{shafter_1986},  and is called the diagnostic diagram. To apply it to our dataset, we fitted each spectrum with a model composed of two Gaussians, expressed in terms of the following parameters: the peak-to-peak separation $a$ (for which we explored the range   $1600-4500\, {\rm km\,s^{-1}}$, in steps of $100\, {\rm km\,s^{-1}}$), the FWHM and height of both Gaussians (which we kept linked), and the offset of the full structure.  We subsequently performed a sinusoidal fit to the resulting radial velocities and determined orbital parameters for each value independently.

Assuming that the contamination is dominant in the external regions, for sufficiently large $a$ values in our modelling we would fit only the wings of the profile, and the measured radial velocities should tend to a stable more reliable measurement that traces the compact object's orbit (Fig. \ref{prova_diagnositc}). Therefore, it is useful to plot the orbital parameters, such as the radial velocity semi-amplitude $K_\mathrm{1}$, the ratio $\sigma_{k}$/$K_\mathrm{1}$ (which estimates the relative error in $K_\mathrm{1}$), the orbital phase (defined under the same convention previously described), and the systemic velocity $\gamma$, against the $a$ parameter.
To determine the maximum reasonable value of $a$, we have to look into how $\sigma_{k}$/$K_\mathrm{1}$ varies with the separation: an increase over the average value implies that the continuum noise is dominant. This technique can therefore provide us with estimates of the orbital parameters, although their value is strongly influenced by the choice of separation $a$, which is somewhat subjective.

As we can see in Fig. \ref{diagnostic}, for $a \lesssim 3800$ $\rm{km\, s^{-1}}$ $K_\mathrm{1}$ remains stable over a range with a mean value and standard deviation of $91 \pm 18 \, \rm{km\,s^{-1}}$;   above this separation the ratio $\sigma_{k}/K_\mathrm{1}$  starts to increase. 
In particular, $K_\mathrm{1}$ does not considerably deviate from the mean value even though we adopt a peak-to-separation in the range $3400-4000\,\rm{km\,s^{-1}}$.  
Additionally, the phase shift, defined as the difference between the value obtained with the method described above and that derived from the double-Gaussian method, which should remain around zero, starts to decrease significantly above $a=3800 \, \rm{km\,s^{-1}}$. These considerations led us to think that the most reliable values of the orbital parameters can be obtained for a separation smaller than $3800 \,\rm{km\, s^{-1}}$, in line with the value obtained by \cite{Corral_santana_2013} based on their analysis of H$\alpha$ ($a=3600\, \rm{km\,s^{-1}}$). Therefore, the best estimate of the radial semi-amplitude velocity would be provided by its average value below the maximum separation (i.e. $K_\mathrm{1} =92 \pm 18 \, \rm{km\, s^{-1}}$), while the systemic velocity of the system remains stable for all the separations with a mean value of $\gamma= -228 \pm 13 \, \rm{km\,s^{-1}}$.

We subsequently corrected each spectrum from its measured velocity to calculate an average spectrum in the reference frame of the binary. 
We performed a fit to this averaged profile by making use of a single-Gaussian profile to obtain the FWHM of the line. The model returns $ \rm{FWHM} = 4574 \pm 156 \, \rm{km\, s^{-1}}$, notably higher than (but still consistent within $\sim 2$ $\sigma$ with)  the value measured in quiescence for H$\alpha$ by \cite{Torres_2015}, \cite{Casares_2015}, and \cite{Daniel_2015}, being $4025 \pm 110 \, \rm{km\,s^{-1}}$, $4085  \pm  328 \, \rm{km\,s^{-1}}$, and  $4152 \pm  209 \, \rm{km\,s^{-1}}$, respectively. 

\section{Discussion}\label{sec_4}
J1357 is a rather unique BH transient with observational properties that are not properly understood, such as recurrent optical dips, presumably as a result of a very high orbital inclination \citep{Corral_santana_2013}, as well as high and variable systemic velocity estimations \citep{Daniel_2015}.
A crucial way to better understand this object is to constrain its fundamental parameters, such as the systemic velocity (key to understanding the system evolution; see e.g. \citealt{Repetto_2015}) and its orbital period. Given that the disc (i.e. not the companion) dominates the optical emission even during the quiescent phase, the study of the emission lines during this stage arguably offers the best opportunity to make progress in this area.

We present the phase-resolved spectroscopy of the system, focused on the $\rm H\beta$ region. The data were obtained deep into the quiescent phase ($\sim$5 yr from the previous outburst) and represent the first study of this line for J1357, and one of the few available for a quiescent LMXB in this spectral region.

We observe a periodic modulation in the radial velocity of the H$\beta$ line profile at $\rm{P}= 0.127 \pm 0.008 \,d$. This result is consistent with that previously found in outburst \citep{Corral_santana_2013} and quiescence \citep{Daniel_2015} based on H$\alpha$ studies. Therefore, we infer that this modulation has remained stable over the five years, further strengthening its association with the orbital period of the system.

We also derive a systemic velocity of $-228 \pm 13 \,\rm{km\,s^{-1}}$ and a K$_\mathrm{1}$ of $91\pm 18\, \rm{km\,s^{-1}}$. 
\cite{Daniel_2015} derived $\gamma=-79 \pm 12 \,\rm{km\,s^{-1}}$, a measurement consistent with another quiescence measurement ($\gamma=-130 \pm 50 \,\rm{km\,s^{-1}}$; \citealt{Torres_2015}). 
Although our result, obtained from an analogous analysis but on the H$\beta$ line, is consistent with \citep{Torres_2015} at the 2$\sigma$ level, it shows a clear deviation from any of the other previously reported results, favouring $\gamma=-228 \pm 13 \,\rm{km\,s^{-1}}$. Therefore, we conclude that the variable systemic velocity of J1357 is not due to an instrumental issue as there are now a number of datasets in quiescence obtained over $\sim$ 3 yr with different instruments and telescopes, and based on the analysis of different lines, showing different results. Therefore, it strengthens the interpretation that the variable systemic velocity is a real property of the system, which we associate with the precession of the accretion disc following \cite{Torres_2015}. The fact that these effects remain after five years of quiescence also shows that this is most likely a persistent property of the system.

Our measurement of the quiescence FHWM of the average H$\beta$ profile is slightly higher than that reported for the H$\alpha$ line in previous works (though still consistent within $2\sigma$; see Sect. \ref{diagnostic}). 
However, it is known that higher-order Balmer lines are typically broader than H$\alpha$ (see \citealt{Marsh_1987,1994_Marsh}). The more energetic H$\beta$ line is expected to be emitted from hotter (i.e. inner) regions of the accretion disc than H$\alpha$, where the velocities are higher, and therefore producing a broader profile. Nevertheless, the consistency of the FWHM of $\rm H\beta$ with that reported for $\rm H\alpha$ two years earlier allows us to infer that the system quiescent properties remained reasonably stable over a period of $\sim$ 2 yr. This demonstrates that J1357 was already in a true quiescent state by then. Together with the orbital period, which we also independently confirm, these are the two key assumptions required for the application of the FWHM--K2 correlation \citep{Casares_2015}. Therefore, we provide strong arguments to support the mass function derived by \citep{Daniel_2015}, which implies one of the most massive BHs found in the Galaxy.

\subsection{Radial velocity  of the compact object}
Measurement of the compact object's radial velocity through the diagnostic diagram is an indirect method that actually traces the accretion disc's innermost optically emitting radii. 
This means that the mismatch of the K$_{1}$ semi-amplitude velocity with the previous measurements in quiescence reported in the literature may open up a point of reflection on the state of the source.

Different works in the literature have investigated the spectral evolution of several transient sources in quiescence after the outburst. 
One of the most recent examples is provided in \cite{Casares_2015}, which analysed the evolution of the FWHM of the H$\alpha$ in the BH LMXB V404 Cygni during the  20 years since the outburst peak. They infer that the FWHM started to increase during the outburst decay, and that it reached a stable plateau phase b (a true quiescence), starting $\sim 1300$ days after the peak of the outburst. 
On the other hand, \cite{2019_russel} detected a brightening at optical wavelengths prior to an X-ray outburst in four X-ray binaries that were leaving their quiescent state. In their work the authors report that the optical precursor occurred weeks before the first X-ray flare, a delay consistent with the time of refilling of the disc that causes the X-ray brightening.

Our observations during the quiescent state of J1357 lie between two outburst events (2011 and 2017), five years after the previous discovery outburst, and one year before the following event. On the other hand, \cite{Daniel_2015} and \cite{Torres_2015} used data from the quiescent epochs between April 2013 and from April to June 2014, respectively (only 2-3 yr after the first outburst). 
It is true that, as mentioned above, \cite{Casares_2015} deduced a period of about 3.5 yr after the outburst for a full quiescent phase to be reached. However, the source analysed by the authors possesses a much larger accretion disc than J1357, so we would expect that our source will take less time to relax in quiescence. 
\cite{Daniel_2015} found $K_\mathrm{1}=40 \pm 12 \, \rm{km\,s^{-1}}$, fully in line with that reported in outburst by \cite{Corral_santana_2013}. Similarly, the parameters presented in \cite{Torres_2015} are in
line with these works. In the present paper, however, the resulting $K_\mathrm{1}$ value is notably larger than those previously derived. The time elapsed between outbursts and quiescent phases studied both in this and previous works does not suggest a different state of the system, and therefore does not explain the different $K_\mathrm{1}$ values.
 
Another explanation could be related to the fact that our analysis is focused on H$\beta$ and not on H$\alpha$. Although both are transitions of the hydrogen Balmer series, the higher energy associated with H$\beta$ might naively suggest that it was formed at an accretion disc ring slightly closer to the compact object than H$\alpha$.
The diagnostic diagram technique has been successfully applied to a number of LMXBs (e.g. \citealt{2013_Cornelisse}, \citealt{2012_Somero}, \citealt{2009_teo}). Nevertheless, systems where $K_\mathrm{1}$ can be independently measured are scarce, and they have shown that the diagnostic diagram results are better understood as an upper limit to the true compact object radial velocity. In this regard, \cite{2006_Sandaza} showed how the diagnostic diagram method may fail to provide a strong determination of $K_\mathrm{1}$, identifying the root cause of this problem with an asymmetry due to a hot spot that extends to higher velocities, up to the wings of the emission line profile. The fact that applying the same technique in a different line and epoch produces a distinct and larger $K_\mathrm{1}$ value than previous works suggests a similar scenario for J1357.

\subsection{Inclination angle\label{estimation_inclination}}
Several observational features suggest a high orbital inclination for J1357 \citep{Corral_santana_2013,Daniel_2015,Armas_2014}.  Arguably, the most debated is the presence of optical dips without X-ray counterparts, which sparked the discussion on the geometry of the source (see Sect. \ref{sec_1}).
Recently, \cite{Casares_2022} has found a linear correlation between the depth of the inner core of the double-peaked H$\alpha$ emission profile and the inclination angle in a set of quiescent X-ray transients. The authors applied this correlation to J1357, deducing an inclination of $87.4^{+2.6}_{-5.6}$ degrees, thus supporting the near edge-on geometry of the source.

Visual inspection of the individual spectra reveals that the core of the H$\beta$ line exhibits narrow and variable absorption components, changing in depth during the orbit and even approaching the continuum level (see Fig. \ref{fig:2fit}).
Deep and narrow absorption cores have been observed in high-inclination cataclysmic variables (with $i\, \ge \ang{75}$ \citealt{1983_Schoembs}), but never on a LMXB, and they are thought to be caused by occultation of the inner regions of the accretion disc.

Hints of the presence of a narrow core on the HeI 5876 line of J1357 were found by \cite{Daniel_2015}, but further confirmation was required due to the low resolution of their spectra, which hampered a clean subtraction of the background. Focusing on the H$\beta$ wavelength range and using the higher-resolution grism, we can now confirm the presence of this feature in a different emission line (H$\beta$), as well as its narrowness and variable nature along the orbit. 
This variability can be used to track the motion of the structure responsible for the occultation, as dips are thought to be caused by structures close to the outer rim of the disc \citep{King}. In particular, the `bulge' is produced by the impact of the stream of matter from the donor star on the disc. As such, it is expected to impact the disc between orbital phases $\sim$ 0.8 and  0, depending on mass ratio \citep{1970_smak}. In order to test this hypothesis, we adopted the absolute phase 0 derived by \cite{Casares_2022} (T0=2456396.6617 HJD, derived from a study of the line through variability in BHs) and redefined the orbital phases for each epoch using the equation mentioned in section \ref{radial velocity}.
Since the narrow core present in our spectra is particularly prominent in spec-1 ($\phi=0.89$), we conclude that this scenario is plausible. However, we note that shallower narrow cores are also present at other orbital phases (e.g. spec-7), suggesting a more complex outer rim geometry.

To further inspect the orbital variability of the profiles, we followed the  \citet{Casares_2022} prescription to calculate the normalised depth of the central trough (T) for all spectra, performing a double symmetric Gaussian fit with equal FWHM and intensity. The variations in T during the orbit appear consistent with the evolution presented in the above-mentioned paper (based on the analysis of \ha{}), and provide further support to the presence of variable inner cores, reaching maximum depth for spec-1 (see Table \ref{2gaussfit}). For this epoch, the best-fit model remains unable to fully fit the observed spectrum core (which almost reaches the continuum level), further supporting the presence of an additional narrow core component. The T value measured in our mean spectrum ($0.585 \pm 0.008$) is consistent with that found by \citet{Casares_2022} ($0.612\pm0.125$), and therefore with the high inclination they deduced from the T-I plane.
We also considered the possibility of a grazing eclipse by the companion star being at the core of this phenomenon. However, the fact that spec-2, obtained at the epoch of inferior conjunction, does not exhibit a  clear deep core led us to disregard this possibility. \cite{Corral_santana_2013} proposed that the extreme mass ratio of the system ($q \sim 0.04$, \citealt{Daniel_2015}) makes the donor star radius comparable to or smaller than the disc outer rim. Thus, the companion would not be expected to produce clear eclipsing phenomena, even in a close-to-edge-on configuration.

\section{Conclusions}
We have analysed observations performed with the GTC telescope of J1357 during the quiescent phase, $\sim$ 5 yr after the previous outburst. Taking advantage of the low extinction of the source, we have performed, for the first time, a phase-resolved study on the H$\beta$ emission line.

We find a periodicity ($\rm{P}= 0.127 \pm 0.008\, \rm{d}$), which is fully consistent with previous reports based on different lines and obtained at different epochs and with different facilities. This strongly suggests that this corresponds to the orbital period of the system, confirming J1357 as one of the shortest orbital period BH transients. 

Our obtained values for systemic velocity and radial semi-amplitude velocity differ from the previous values derived from the H$\alpha$ emission line. We interpret this as an effect of a combination of disc precession and the presence of hot spots. 
Finally, we report, for the first time, the presence of narrow, deep, and variable cores in the H$\beta$ emission line of a LMXB. We propose that these cores are produced by structures in the outer ring of the accretion disc, which we recognise as the outer disc bulge. This provides independent proof to further support the high-inclination nature of the system.

\begin{acknowledgements}
The authors acknowledge financial contribution from the agreement ASI-INAF n.2017-14-H.0 and INAF mainstream (PI: A. De Rosa, T. Belloni), from the HERMES project financed by the Italian Space Agency (ASI) Agreement n. 2016/13 U.O and from the ASI-INAF Accordo Attuativo HERMES Technologic Pathfinder n. 2018-10-H.1-2020. We also acknowledge support from the European Union Horizon 2020 Research and Innovation Framework Programme under grant agreement HERMES-Scientific Pathfinder n. 821896 and from PRIN-INAF 2019 with the project "Probing the geometry of accretion: from theory to observations" (PI: Belloni). DMS acknowledges support from the Consejería de Economía, Conocimiento y Empleo del Gobierno de Canarias and the European Regional Development Fund (ERDF) under a grant with reference ProID2021010132 (ACCISI/FEDER, UE); as well as from the Spanish Ministry of Science and Innovation via a Europa Excelencia grant (EUR2021-122010) and grant PID2020-120323GB-I00.

\end{acknowledgements}
\bibliography{46909corr}

\begin{thebibliography}{32}
\expandafter\ifx\csname natexlab\endcsname\relax\def\natexlab#1{#1}\fi

\bibitem[{{Anitra} {et~al.}(2021){Anitra}, {Di Salvo}, {Iaria}, {Burderi}, {Gambino}, {Mazzola}, {Marino}, {Sanna}, \& {Riggio}}]{anitra_2021}
{Anitra}, A., {Di Salvo}, T., {Iaria}, R., {et~al.} 2021, \aap, 654, A160

\bibitem[{{Armas Padilla} {et~al.}(2013){Armas Padilla}, {Degenaar}, {Russell}, \& {Wijnands}}]{monse_2013}
{Armas Padilla}, M., {Degenaar}, N., {Russell}, D.~M., \& {Wijnands}, R. 2013, \mnras, 428, 3083

\bibitem[{{Armas Padilla} {et~al.}(2014{\natexlab{a}}){Armas Padilla}, {Wijnands}, {Altamirano}, {M{\'e}ndez}, {Miller}, \& {Degenaar}}]{Armas_2014}
{Armas Padilla}, M., {Wijnands}, R., {Altamirano}, D., {et~al.} 2014{\natexlab{a}}, \mnras, 439, 3908

\bibitem[{{Armas Padilla} {et~al.}(2014{\natexlab{b}}){Armas Padilla}, {Wijnands}, {Degenaar}, {Mu{\~n}oz-Darias}, {Casares}, \& {Fender}}]{2014_armasB}
{Armas Padilla}, M., {Wijnands}, R., {Degenaar}, N., {et~al.} 2014{\natexlab{b}}, \mnras, 444, 902

\bibitem[{{Astropy Collaboration} {et~al.}(2022){Astropy Collaboration}, {Price-Whelan}, {Lim}, {Earl}, {Starkman}, {Bradley}, {Shupe}, {Patil}, {Corrales}, {Brasseur}, {N{\"o}the}, {Donath}, {Tollerud}, {Morris}, {Ginsburg}, {Vaher}, {Weaver}, {Tocknell}, {Jamieson}, {van Kerkwijk}, {Robitaille}, {Merry}, {Bachetti}, {G{\"u}nther}, {Aldcroft}, {Alvarado-Montes}, {Archibald}, {B{\'o}di}, {Bapat}, {Barentsen}, {Baz{\'a}n}, {Biswas}, {Boquien}, {Burke}, {Cara}, {Cara}, {Conroy}, {Conseil}, {Craig}, {Cross}, {Cruz}, {D'Eugenio}, {Dencheva}, {Devillepoix}, {Dietrich}, {Eigenbrot}, {Erben}, {Ferreira}, {Foreman-Mackey}, {Fox}, {Freij}, {Garg}, {Geda}, {Glattly}, {Gondhalekar}, {Gordon}, {Grant}, {Greenfield}, {Groener}, {Guest}, {Gurovich}, {Handberg}, {Hart}, {Hatfield-Dodds}, {Homeier}, {Hosseinzadeh}, {Jenness}, {Jones}, {Joseph}, {Kalmbach}, {Karamehmetoglu}, {Ka{\l}uszy{\'n}ski}, {Kelley}, {Kern}, {Kerzendorf}, {Koch}, {Kulumani}, {Lee}, {Ly}, {Ma}, {MacBride}, {Maljaars}, {Muna}, {Murphy}, {Norman},
  {O'Steen}, {Oman}, {Pacifici}, {Pascual}, {Pascual-Granado}, {Patil}, {Perren}, {Pickering}, {Rastogi}, {Roulston}, {Ryan}, {Rykoff}, {Sabater}, {Sakurikar}, {Salgado}, {Sanghi}, {Saunders}, {Savchenko}, {Schwardt}, {Seifert-Eckert}, {Shih}, {Jain}, {Shukla}, {Sick}, {Simpson}, {Singanamalla}, {Singer}, {Singhal}, {Sinha}, {Sip{\H{o}}cz}, {Spitler}, {Stansby}, {Streicher}, {{\v{S}}umak}, {Swinbank}, {Taranu}, {Tewary}, {Tremblay}, {de Val-Borro}, {Van Kooten}, {Vasovi{\'c}}, {Verma}, {de Miranda Cardoso}, {Williams}, {Wilson}, {Winkel}, {Wood-Vasey}, {Xue}, {Yoachim}, {Zhang}, {Zonca}, \& {Astropy Project Contributors}}]{astropy}
{Astropy Collaboration}, {Price-Whelan}, A.~M., {Lim}, P.~L., {et~al.} 2022, \apj, 935, 167

\bibitem[{{Casares}(2015)}]{Casares_2015}
{Casares}, J. 2015, \apj, 808, 80

\bibitem[{{Casares} {et~al.}(2022){Casares}, {Mu{\~n}oz-Darias}, {Torres}, {Mata S{\'a}nchez}, {Britt}, {Armas Padilla}, {{\'A}lvarez-Hern{\'a}ndez}, {C{\'u}neo}, {Gonz{\'a}lez Hern{\'a}ndez}, {Jim{\'e}nez-Ibarra}, {Jonker}, {Panizo-Espinar}, {S{\'a}nchez-Sierras}, \& {Yanes-Rizo}}]{Casares_2022}
{Casares}, J., {Mu{\~n}oz-Darias}, T., {Torres}, M.~A.~P., {et~al.} 2022, \mnras, 516, 2023

\bibitem[{{Charles} {et~al.}(2019){Charles}, {Matthews}, {Buckley}, {Gandhi}, {Kotze}, \& {Paice}}]{Charles_2019}
{Charles}, P., {Matthews}, J.~H., {Buckley}, D. A.~H., {et~al.} 2019, \mnras, 489, L47

\bibitem[{{Cornelisse} {et~al.}(2013){Cornelisse}, {Kotze}, {Casares}, {Charles}, \& {Hakala}}]{2013_Cornelisse}
{Cornelisse}, R., {Kotze}, M.~M., {Casares}, J., {Charles}, P.~A., \& {Hakala}, P.~J. 2013, \mnras, 436, 910

\bibitem[{{Corral-Santana} {et~al.}(2013){Corral-Santana}, {Casares}, {Shahbaz}, {Zurita}, {Mart{\'\i}nez-Pais}, \& {Rodr{\'\i}guez-Gil}}]{Corral_santana_2013}
{Corral-Santana}, J.~M., {Casares}, J., {Shahbaz}, T., {et~al.} 2013, in Revista Mexicana de Astronomia y Astrofisica Conference Series, Vol.~42, Revista Mexicana de Astronomia y Astrofisica Conference Series, 3--4

\bibitem[{{Frank} {et~al.}(2002){Frank}, {King}, \& {Raine}}]{King}
{Frank}, J., {King}, A., \& {Raine}, D.~J. 2002, {Accretion Power in Astrophysics: Third Edition}

\bibitem[{{Iaria} {et~al.}(2013){Iaria}, {Di Salvo}, {D'A{\`\i}}, {Burderi}, {Mineo}, {Riggio}, {Papitto}, \& {Robba}}]{2013_Iaria}
{Iaria}, R., {Di Salvo}, T., {D'A{\`\i}}, A., {et~al.} 2013, \aap, 549, A33

\bibitem[{{Jim{\'e}nez-Ibarra} {et~al.}(2019){Jim{\'e}nez-Ibarra}, {Mu{\~n}oz-Darias}, {Casares}, {Armas Padilla}, \& {Corral-Santana}}]{2019_jimenez-ibarra}
{Jim{\'e}nez-Ibarra}, F., {Mu{\~n}oz-Darias}, T., {Casares}, J., {Armas Padilla}, M., \& {Corral-Santana}, J.~M. 2019, \mnras, 489, 3420

\bibitem[{{Kennea} {et~al.}(2012){Kennea}, {Yang}, {Altamirano}, {Evans}, {Krimm}, {Romano}, {Mangano}, {Curran}, {Yamaoka}, \& {Negoro}}]{2012_Kennea}
{Kennea}, J.~A., {Yang}, Y.~J., {Altamirano}, D., {et~al.} 2012, The Astronomer's Telegram, 4044, 1

\bibitem[{{Kuulkers} {et~al.}(2013){Kuulkers}, {Kouveliotou}, {Belloni}, {Cadolle Bel}, {Chenevez}, {D{\'\i}az Trigo}, {Homan}, {Ibarra}, {Kennea}, {Mu{\~n}oz-Darias}, {Ness}, {Parmar}, {Pollock}, {van den Heuvel}, \& {van der Horst}}]{2013_Kuulkers}
{Kuulkers}, E., {Kouveliotou}, C., {Belloni}, T., {et~al.} 2013, \aap, 552, A32

\bibitem[{{Marsh} {et~al.}(1987){Marsh}, {Horne}, \& {Shipman}}]{Marsh_1987}
{Marsh}, T.~R., {Horne}, K., \& {Shipman}, H.~L. 1987, \mnras, 225, 551

\bibitem[{{Marsh} {et~al.}(1994){Marsh}, {Robinson}, \& {Wood}}]{1994_Marsh}
{Marsh}, T.~R., {Robinson}, E.~L., \& {Wood}, J.~H. 1994, \mnras, 266, 137

\bibitem[{{Mata S{\'a}nchez} {et~al.}(2015){Mata S{\'a}nchez}, {Mu{\~n}oz-Darias}, {Casares}, {Corral-Santana}, \& {Shahbaz}}]{Daniel_2015}
{Mata S{\'a}nchez}, D., {Mu{\~n}oz-Darias}, T., {Casares}, J., {Corral-Santana}, J.~M., \& {Shahbaz}, T. 2015, \mnras, 454, 2199

\bibitem[{{Mu{\~n}oz-Darias} {et~al.}(2009){Mu{\~n}oz-Darias}, {Casares}, {O'Brien}, {Steeghs}, {Mart{\'\i}nez-Pais}, {Cornelisse}, \& {Charles}}]{2009_teo}
{Mu{\~n}oz-Darias}, T., {Casares}, J., {O'Brien}, K., {et~al.} 2009, \mnras, 394, L136

\bibitem[{{Naylor}(1998)}]{1998_Naylor}
{Naylor}, T. 1998, \mnras, 296, 339

\bibitem[{{Paice} {et~al.}(2019){Paice}, {Gandhi}, {Charles}, {Dhillon}, {Marsh}, {Buckley}, {Kotze}, {Beri}, {Altamirano}, {Middleton}, {Plotkin}, {Miller-Jones}, {Russell}, {Tomsick}, {D{\'\i}az-Merced}, \& {Misra}}]{paice_2019}
{Paice}, J.~A., {Gandhi}, P., {Charles}, P.~A., {et~al.} 2019, \mnras, 488, 512

\bibitem[{{Repetto} \& {Nelemans}(2015)}]{Repetto_2015}
{Repetto}, S. \& {Nelemans}, G. 2015, \mnras, 453, 3341

\bibitem[{{Russell} {et~al.}(2019){Russell}, {Bramich}, {Lewis}, {AlMannaei}, {Al Qaissieh}, {Al Qasim}, {Al Yazeedi}, {Baglio}, {Bernardini}, {Elgalad}, {Gabuya}, {Lasota}, {Palado}, {Roche}, {Shivkumar}, {Udrescu}, \& {Zhang}}]{2019_russel}
{Russell}, D.~M., {Bramich}, D.~M., {Lewis}, F., {et~al.} 2019, Astronomische Nachrichten, 340, 278

\bibitem[{{Schoembs} \& {Hartmann}(1983)}]{1983_Schoembs}
{Schoembs}, R. \& {Hartmann}, K. 1983, \aap, 128, 37

\bibitem[{{Shafter} {et~al.}(1986){Shafter}, {Szkody}, \& {Thorstensen}}]{shafter_1986}
{Shafter}, A.~W., {Szkody}, P., \& {Thorstensen}, J.~R. 1986, \apj, 308, 765

\bibitem[{{Smak}(1970)}]{1970_smak}
{Smak}, J. 1970, \actaa, 20, 311

\bibitem[{{Somero} {et~al.}(2012){Somero}, {Hakala}, {Muhli}, {Charles}, \& {Vilhu}}]{2012_Somero}
{Somero}, A., {Hakala}, P., {Muhli}, P., {Charles}, P., \& {Vilhu}, O. 2012, \aap, 539, A111

\bibitem[{{Torres} {et~al.}(2015){Torres}, {Jonker}, {Miller-Jones}, {Steeghs}, {Repetto}, \& {Wu}}]{Torres_2015}
{Torres}, M.~A.~P., {Jonker}, P.~G., {Miller-Jones}, J.~C.~A., {et~al.} 2015, \mnras, 450, 4292

\bibitem[{{Unda-Sanzana} {et~al.}(2006){Unda-Sanzana}, {Marsh}, \& {Morales-Rueda}}]{2006_Sandaza}
{Unda-Sanzana}, E., {Marsh}, T.~R., \& {Morales-Rueda}, L. 2006, \mnras, 369, 805

\bibitem[{{van Dokkum} {et~al.}(2012){van Dokkum}, {Bloom}, \& {Tewes}}]{2012_lacosmic}
{van Dokkum}, P.~G., {Bloom}, J., \& {Tewes}, M. 2012, {L.A.Cosmic: Laplacian Cosmic Ray Identification}, Astrophysics Source Code Library, record ascl:1207.005

\bibitem[{Virtanen {et~al.}(2020)Virtanen, Gommers, Oliphant, Haberland, Reddy, Cournapeau, Burovski, Peterson, Weckesser, Bright, {van der Walt}, Brett, Wilson, Millman, Mayorov, Nelson, Jones, Kern, Larson, Carey, Polat, Feng, Moore, {VanderPlas}, Laxalde, Perktold, Cimrman, Henriksen, Quintero, Harris, Archibald, Ribeiro, Pedregosa, {van Mulbregt}, \& {SciPy 1.0 Contributors}}]{2020SciPy-NMeth}
Virtanen, P., Gommers, R., Oliphant, T.~E., {et~al.} 2020, Nature Methods, 17, 261

\bibitem[{{Wijnands} {et~al.}(2006){Wijnands}, {in't Zand}, {Rupen}, {Maccarone}, {Homan}, {Cornelisse}, {Fender}, {Grindlay}, {van der Klis}, {Kuulkers}, {Markwardt}, {Miller-Jones}, \& {Wang}}]{2006_Wijnands}
{Wijnands}, R., {in't Zand}, J.~J.~M., {Rupen}, M., {et~al.} 2006, \aap, 449, 1117

\end{thebibliography}
\bibliographystyle{aa}

\end{document}